\documentclass{article}
\usepackage{spconf,amsmath,graphicx}
\usepackage{booktabs}
\usepackage{amssymb}
\usepackage{url}
\usepackage[subtle, tracking=normal]{savetrees}
\usepackage{multirow}
\usepackage[export]{adjustbox}

\title{The DKU Post-Challenge Audio-Visual Wake Word Spotting System \\
for the 2021 MISP Challenge: Deep Analysis}
\name{Haoxu Wang$^{1,2}$, Ming Cheng$^{1,2}$, Qiang Fu$^{3}$, Ming Li$^{1,2\dagger}$\thanks{$\dagger$ Corresponding Author, E-mail: 	ming.li369@dukekunshan.edu.cn}}
\address{$^{1}$School of Computer Science, Wuhan University, Wuhan, China \\
$^{2}$Data Science Research Center, Duke Kunshan University, Kunshan, China \\
$^{3}$Alibaba Group, China \\
}

\begin{document}
\ninept

\maketitle
%



\begin{abstract}
This paper further explores our previous wake word spotting system ranked 2-nd in Track 1 of the MISP Challenge 2021. First, we investigate a robust unimodal approach based on 3D and 2D convolution and adopt the simple attention module (SimAM) for our system to improve performance. Second, we explore different combinations of data augmentation methods for better performance. Finally, we study the fusion strategies, including score-level, cascaded and neural fusion. Our proposed multimodal system leverages multimodal features and uses the complementary visual information to mitigate the performance degradation of audio-only systems in complex acoustic scenarios. Our system obtains a false reject rate of 2.15\% and a false alarm rate of 3.44\% in the evaluation set of the competition database, which achieves the new state-of-the-art performance by 21\% relative improvement compared to previous systems. Related resource can be found at: \url{https://github.com/Mashiro009/DKU_WWS_MISP}.
\end{abstract}

\section{Introduction}
\label{sec:intro}

With the rapid application of hand-free devices such as mobile phones and voice assistants, Wake Word Spotting (WWS), also known as a specific case of Keyword Spotting (KWS), is increasingly attractive. WWS system is designed to detect a predefined wake word or a set of wake words in the streaming audio.

Recently, many works for WWS based on deep neural network are proposed, including deep neural networks (DNN) \cite{Sun2017CompressedTD},  
convolutional neural networks (CNN) \cite{Sainath2015ConvolutionalNN}, temporal convolutional neural networks \cite{TemporalCnn1} and Transformer \cite{kwsTransformer}. These works show good performance under clean and close-talking scenarios. However, it is observed that the probability of the false alarm becomes higher under complex acoustic environments such as background noises (cheers, TV or screams), reverberations, and conversational multi-speaker interactions with a significant portion of speech overlaps, which will harm the user experience in practical applications.


Although there has been a great deal of multi-modal works in the field of audio-visual speech recognition (AVSR) \cite{afouras2018deep,makino2019recurrent}, previous works mainly focus on the audio-only WWS system, which ignores the use of visual information. Inspired by human perception, complementary visual information (e.g., lip movements) can help understand the semantics of speech in complex acoustic scenarios since the visual modal is not affected by acoustic noise. \cite{AVSR_conformer} uses conformer \cite{asr_conformer} architecture in AVSR, and \cite{hong22_interspeech} investigates an audio enhancement module called V-CAFE for AVSR. In \cite{seewakewords}, the authors investigate a simple CNN-based audio-visual KWS system and demonstrate good performance for noisy audio environments. Along with the 1st Multimodal Information based Speech Processing Challenge (MISP Challenge 2021 \cite{mispchallenge}) and its data release \cite{zhou22g_interspeech}, some new works are proposed for Audio-Visual Wake Word Spotting (AVWWS) including CNN-3D-based model \cite{cheng2022dku} and transformer-based end-to-end model \cite{xu2022audio}.

In this paper, we extend our previous work \cite{cheng2022dku} to an effective AVWWS system. 
We design the robust unimodal system by replacing previous  3D CNN with a mixture of 2D and 3D CNNs. 
We also adopt the simple attention module (SimAM) \cite{yang2021simam} to our WWS system without extra parameters, which has been used in automatic speaker verification (ASV) \cite{qin2022simple,xiaoyitaslp}. To further improve the performance of our WWS system, we do experiments to explore better combination of the data augmentation methods. Finally, we study the fusion strategies for the two unimodal systems and obtain a 21\% relative improvement on top of our previous second-place work. Also, we achieve the state-of-the-art (SOTA) result (5.59\% WWS score) on the MISP dataset.

\begin{figure}[t]
    \centering
    \includegraphics[width=0.4\textwidth]{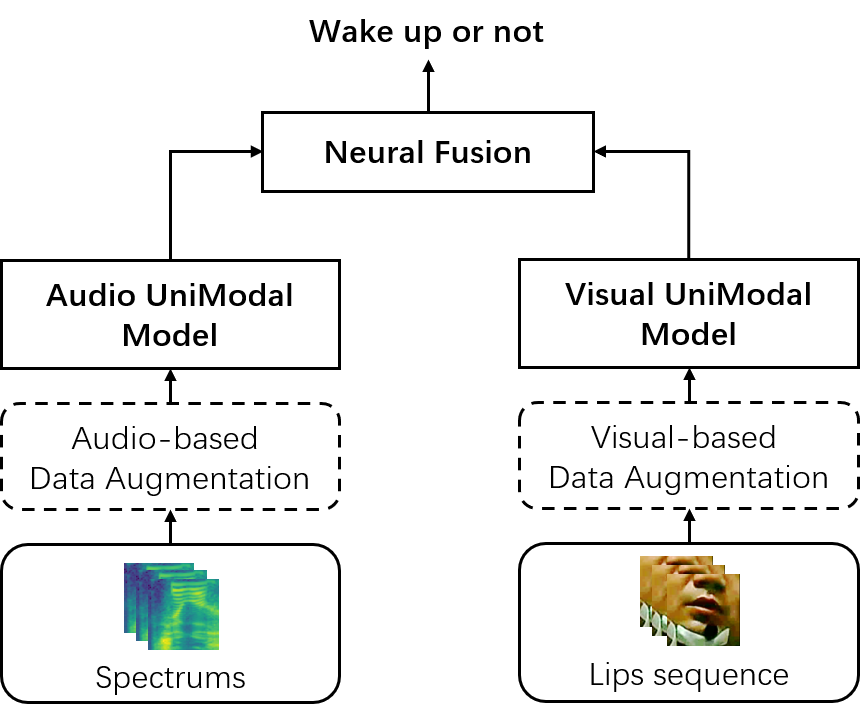}
    \caption{Framework of our audio-visual wake word spotting system.}
    \label{framework}
    \vspace{-0.8em}
\end{figure}

\section{System Description}

\vspace{-0.5em}
As shown in Figure \ref{framework}, we first investigate a robust model for single-modality and then study the fusion strategies for the two solid unimodal systems to get an effective audio-visual WWS system. We also investigate the data augmentation strategies and find a better combination than the baseline. 

\subsection{Unimodal Models}

\subsubsection{3D-ResNet34}

Our previous work \cite{cheng2022dku} constructs a ResNet34-liked neural network based on 3D CNNs (ResNet-3D) to cope with the unimodal WWS task. The two modalities have the same network architecture, except that the dimensions of the inputs are different. The video inputs are raw image sequences with the shape of $(T, H, W, 3)$, in which $T$ represents the number of frames, $H, W$ represents the height and width of the image, and $3$ represents the RGB channel. For the audio inputs, we first extract the 2D acoustic features $X_{spec} \in \mathbb{R}^{T^{'} \times D}$(e.g. FBank, MFCC) from the 1D waveform, and then we use a sliding window with the shape of $(
D, D)$ to slice the 2D acoustic features along the time axis with the stride of $\frac{T'}{T}$ because of the difference of the fps of audio and video, so we organize the audio input $X_{audio}$ with the shape of $(T, D, D, 1)$. Hence, we can make the decision using the unimodal model with inputs that have the same shape of $(T, H, W, C)$.
\vspace{-0.8em}
\subsubsection{2D-ResNet34}

In previous approaches for audio-only WWS tasks \cite{Sainath2015ConvolutionalNN,TemporalCnn1,kwsTransformer}, the 2D spectrum $X_{spec}$ is usually considered as a single channel image, we also design the 2D-ResNet34 network for such input. 
The experiments in \cite{tran2018closer} and Sec. \ref{sec:experiment} show that 3D-ResNet is more suitable for the temporal-spatial features compared with 2D-ResNet, and the constructed $X_{audio}$ contains local short-time spatial information which 3D-ResNet can better utilize. 
\vspace{-0.8em}
\subsubsection{3D-ResNet18+2D-Resnet18}

\cite{tran2018closer} indicates that 3D CNNs at low-level layers focus on short-term spatial modeling, and 2D CNNs help extract semantic and temporal information at high-level layers in the Action Recognition task. Hence, we split the 3D-ResNet34 into two cascaded networks, which are 3D-ResNet18 and 2D-ResNet18. The design of converting 3D CNN to 2D CNN can reduce the number of parameters of the model. At the same time, the network with a mixture of 3D CNN and 2D CNN can collaborate on the modeling, allowing different modules to focus on different perspectives. The inputs of this network are the same as those of the 3D-ResNet34, but the feature map is averaged along the spatial axis after the 3D-ResNet18 to the shape of $(C, T)$. Then we consider this as a one-channel latent image with the shape of $(1, T, C)$ and send it to the 2D-ResNet18.
\vspace{-0.8em}
\subsubsection{3D-ResNet18+2D-Resnet18+SimAM}


The attention modules for CNN have been widely used and have achieved great performance in computer vision and speech processing. Previous channel-wise squeeze-excitation (SE) \cite{hu2018squeeze} obtains 1D channel-level weights and has been used in WWS \cite{sekws}. Convolutional block attention module (CBAM) calculates 2D and 1D attention weights separately \cite{woo2018cbam}. All of these attention modules need extra parameters.

\cite{yang2021simam} proposes SimAM based on the phenomenon of spatial suppression \cite{webb2005early} in neuroscience. The 3D weights estimate the importance of individual neurons, and spatial suppression shows that the most informative neurons usually show distinctive firing patterns from surrounding neurons. So the energy function for each target neuron output $t$ in the 3D feature map $X \in \mathbb{R}^{C \times H \times W}$ is defined as:

\begin{equation}
\label{eq1}
\begin{array}{cc}
e_{t}(w_{t}, b_{t}, y, x_{i}) = (y_{t} - \hat{t})^{2} + \frac{1}{M-1}\sum^{M-1}_{i=1}{(y_{o} - \hat{x}_{i})^{2}}
\end{array}
\end{equation}

\noindent where $\hat{t} = w_{t}t + b_{t}$ and $\hat{x}_{i} = w_{t}x_{i} + b_{t}$ are the linear transforms of $t$ and $x_{i}$, $x_{i}$ represents the output of neurons other than $t$ in the same channel of the feature map. $i$ is the index along the spatial dimension $M = H \times W$. The energy function attains the minimal value when $\hat{t} = y_{t}$ and $\hat{x}_{i} = y_{o}$, and $y_{t}$ should be larger than $y_{o}$ since the spatial suppression. For simplicity, \cite{yang2021simam} uses the binary label to set the $y_{t} = 1$ and $y_{o} = -1$, and also adds a regularizer into Equation \ref{eq1}:

\begin{equation}
\label{eq2}
\begin{array}{cc}
e_{t}(w_{t}, b_{t}, y, x_{i}) = \frac{1}{M-1}\sum^{M-1}_{i=1}{(-1 - (w_{t}x_{i} + b_{t}))^{2}} + \\ (1 - (w_{t}t + b_{t}))^{2}  + \lambda w^{2}_{t}
\end{array}
\end{equation}

It is computationally expensive to solve the above function. Luckily, Equation \ref{eq2} has a fast closed-form solution with respect to $w_{t}$ and $b_{t}$ according to \cite{yang2021simam}. Then we can get the final minimal energy by putting $w_{t}$ and $b_{t}$ back to equation \ref{eq2}:

\begin{equation}
\begin{array}{c}
e_{t}^{*} = \frac{4(\hat{\sigma}^{2} + \lambda)}{(t-\hat{\mu})^{2} + 2\hat{\sigma}^{2} + 2\lambda}

\end{array}
\end{equation}

\noindent where $\hat{\mu} = \frac{1}{M}\sum^{M}_{i=1}{x_{i}}$ and $\hat{\sigma}^{2} = \frac{1}{M}\sum^{M}_{i=1}{(x_{i} - \hat{\mu}^{2})}$. The importance of each neuron can be obtained by $\frac{1}{e_{t}^{*}}$ based on the spatial suppression. The final SimAM is defined as:

\begin{equation}
\begin{array}{cc}

\widetilde{X} = \sigma(\frac{1}{E}) \bigotimes X

\end{array}
\end{equation}

\noindent where $\bigotimes$ represents the element-wise multiplication, 
$E$ groups all $e_{t}^{*}$
across channel and spatial dimensions for the 3D feature map X,
and $\sigma(\cdot)$ represents the sigmoid function. SimAM calculates the 3D attention weights based on the energy function without extra parameters. Moreover, it is easy to apply the SimAM to the 4D feature map $X_{\mbox{4D}} \in \mathbb{R}^{C \times T \times H \times W}$ of the 3D CNN, we calculate the energy value for each neuron along the temporal and spatial dimension and set $M = T \times H \times W$. 


\subsection{Fusion Strategy}

\subsubsection{Score-Level Fusion}

To fusing the audio-visual systems together, the official baseline makes the final decision just by  weighted summation from the separate unimodal system. 

\begin{equation}
\begin{array}{cc}

P_{av} = \alpha \times P_{a} + \beta \times P_{v}

\end{array}
\end{equation}

\noindent where $P_{a}$ and $P_{v}$ are the posterior probabilities generated by the audio- and video-only systems. However, the posterior probability distributions of the audio- and video-only systems are pretty different. Hence, we need to investigate a more suitable fusion strategy. 

\subsubsection{Cascaded Fusion}

Since there is a performance gap between the two systems, we can use a cascade scheme to let one system with poorer performance detect the wake-up words samples with low confidence $th_{l}$ first and then let the other system with better performance give the final result with a higher threshold $th_{h}$. Here we choose the video-only system as the first stage and the audio-only system as the second stage.

\subsubsection{Neural Fusion}

The latent features extracted from the unimodal system are rich in terms of semantic information and can be used to determine whether they are wake-up word samples. Hence, we train a simple neural network to use these latent features further. 


We use the Hierarchical Modality Aggregation (HMA) approach in our previous work \cite{cheng2022dku}. We apply global average pooling to the latent feature maps from the hierarchical residual CNN blocks in each unimodal neural network at different layer levels. We concatenate the obtained embedding vectors at the same level from two modalities to get multimodal embedding vectors $\mathbf{c}_{l}$, where $l$ represents layer level. Then we use the HMA to aggregate all the $\mathbf{c}_{l}$ and give the final multimodal results.
\vspace{-0.8em}
\subsection{Data Augmentation}


\subsubsection{Audio Stream}
\label{sec:augsec}

For the audio stream, we have tried some audio-base data augmentation methods according to \cite{hong22_interspeech,cheng2022dku,mispchallenge},  including offline noise /reverberation adding, beamforming enhancement, negative sub-segmentation, speed perturbation, volume perturbation, trimming slightly and SpecAugment \cite{specaugment}. 

\begin{itemize}
    \item Offline noise/reverberation adding (NR): We generate room impulse response to the original near-field data to simulate mid and far-field data by using the pyroomacoustic tool and mix noises provided by the officials with a random SNR from -15 to 15 dB referring to the official codes.
    \item Beamforming enhancement (BE): We implement beamforming (MVDR \cite{mvdr}) to multi-channel audio signals to exhaust the potential of microphone arrays and incorporate these beamforming-enhanced audios into the training data.
    \item Negative Sub-segmentation (NS): It is possible for the model to detect the wake-up word only according to the length of the audio with low accuracy since the duration distribution of the positive and negative samples are quite different. Hence, we refer to \cite{Wang2020} sub-segment the negative samples during training. 
	\item Volume perturbation (VP): Each audio is randomly selected to change its volume. The volume changes in the range $[0.125, 2]$.
	\item Speed perturbation (SP): The speed of the audio is randomly changed to be faster or slower, with the ratio in the range $[0.9,1.1]$. 
	\item Trimming slightly (TS): Each audio is randomly selected to be trimmed slightly at the beginning or end, with the ratio of 0.95.
	\item SpecAugment (SA): we use the frequency masking and the time masking for each randomly selected audio.
\end{itemize}
\vspace{-0.8em}
\subsubsection{Video Stream}

For the visual stream, we adopt multiple video-based data augmentation methods, including speed perturbation, frame-wise rotation, horizontal flip, frame-level cropping, color jitters and gray scaling. Exact details can be found in \cite{cheng2022dku}.

\begin{table*}[!htb]\centering
\footnotesize
    \caption{\label{ab} {\it Performance of different data augmentation methods based on audio-only 2D-ResNet34 for the far-field data. NR, BE, NS, VP, SP, TS and SA represent offline noise/reverberation adding, beamforming enhancement, negative sub-segmentation, volume perturbation, speed perturbation, trimming slightly and specaugment in Sec. \ref{sec:augsec}.
    \vspace{+0.2em}
  }}
    \begin{tabular}{ccccccccccccc}
    \toprule
    \multirow{2}*{\textbf{ID}} & \multicolumn{6}{c}{\textbf{DA}}  & \multicolumn{3}{c}{\textbf{Dev[\%]}} & \multicolumn{3}{c}{\textbf{Eval[\%]}}  \\
    \cmidrule(lr){2-7} \cmidrule(lr){8-10} \cmidrule(lr){11-13}  
     & NR+BE & NS & VP & SP & TS & SA & FRR & FAR & WWS & FRR & FAR & WWS \\
    \midrule


    D1 & - & - & - & - & - & - & 12.5 & 3.42 & 15.92 & 16.62 & 4.75 & 21.36 \\
    D2 & $\checkmark$ & - & - & - & - & - & 10.42 & 5.68 & 16.09 & 13.24 & 6.77 & 20.01 \\
    D3 & $\checkmark$ & $\checkmark$ & - & - & - & - & 9.78 & 5.44 & 15.21 & 12.08 & 8.12 & 20.2 \\
    D4 & $\checkmark$ & $\checkmark$ & $\checkmark$ & - & - & - & 8.33 & 6.35 & 14.68 & 13.12 & 7.42 & 20.54 \\
    D5 & $\checkmark$ & $\checkmark$ & - & $\checkmark$ & - & - & 7.53 & 4.62 & 12.15 & 7.66 & 8.59 & 16.25 \\
    D6 & $\checkmark$ & $\checkmark$ & - & - & $\checkmark$ & - & 8.65 & 4.71 & 13.37 & 10.79 & 6.35 & 17.14 \\
    D7 & $\checkmark$ & $\checkmark$ & - & - & - & $\checkmark$ & 5.45 & 7.12 & 12.57 & 7.05 & 10.15 & 17.2 \\
    
    \midrule
    
    D8 & $\checkmark$ & $\checkmark$ & - & $\checkmark$ & $\checkmark$ & $\checkmark$ & 5.45 & 4.76 & 10.21 & 5.64 & 6.05 &  \bf{11.69}  \\
    D9 & $\checkmark$ & - & - & $\checkmark$ & $\checkmark$ & $\checkmark$ & 6.09 & 4.43 & 10.51 & 4.97 & 7.33 &  12.3  \\

    \bottomrule

		\end{tabular}
		\vspace{-1.5em}
\end{table*}

\section{Experimental Setup}
\vspace{-0.5em}
\subsection{Dataset and Evaluation Metrics}
We use the dataset provided by the MISP Challenge 2021. The target of the challenge is to detect the wake word 'Xiao T, Xiao T' spoken by the participants in the far-field smart-home scenarios.

The released database has two subsets: the training set (47k+ negative samples and 5K+ positive samples) and the development set (dev: 2k+ negative samples and 600+ positive samples), which are in the near, middle and far fields. Moreover, an evaluation set (eval: 8K+) without annotations is provided to competition participants, which is only in the far field. After the competition, we get the annotations from the MISP committee to compare our results with other teams fairly.

In the AVWWS task, the positive class represents the existence of the wake word in a given sample, and the negative class indicates the opposite. Following the requirements of the evaluation plan, we use False Reject Rate (FRR), False Alarm Rate (FAR), and the Score of WWS as the evaluation criteria. Let $N_{wake}$ denote the number of samples that contain the wake word, and $N_{non\_wake}$ represent the number of samples without the wake word. The FRR and FAR are defined as follows:

\begin{equation}
\footnotesize
FRR = \frac{N_{FR}}{N_{wake}}, \quad FAR = \frac{N_{FA}}{N_{non\_wake}}
\end{equation}

\noindent where $N_{FR}$ denotes the number of samples containing the wake word while not recognized by the system. $N_{FA}$ denotes the number of samples containing no wake words while predicted to be positive by the system. Hence, the final score of Wake Word Spotting (WWS) is defined as:

\begin{equation}
\footnotesize
Score^{WWS} = FRR + FAR
\end{equation}

\vspace{-0.8em}
\subsection{Data Preprocess}

For the visual steam, we only use the RGB lip region images of the video. We employ a face detector (RetinaFace \cite{deng2019retinaface}) to get the face images and facial landmarks from the video. And then, we deploy a face recognizer (ArcFace \cite{deng2019arcface}) to select the target speaker in the far-field video. Finally, we crop the lip regions of the target speaker based on the detected facial landmarks. The details can be found in \cite{cheng2022dku}. Each extracted lip-region video is resized to have a resolution of $112\times112$ with 3 RGB channels. The dimension $T$ is set to 64, which means each video is sampled to contain 64 frames. Therefore, the shape of the video sample becomes $(64,112,112,3)$.

For the audio stream, we extract the 80-dim log-mel filterbank features(FBank) from the waveform with 25ms long and 10ms shift. The dimension $T^{'}$ is set to 256, which means each audio is sampled to contain 64 frames. Therefore, the shape of the audio sample becomes $(256,80)$. For 3D-ResNet input, we use a sliding window with the shape of $(80, 80)$ to slice the features with the stride of $4$. Hence, we organize the audio samples with the shape of $(64, 80, 80, 1)$.

\vspace{-0.8em}
\subsection{Model Details}

For SimAM, the hyper-parameter $\lambda$ is set to 0.001. For Score-Level Fusion. $\alpha$ and $\beta$ are set to 0.5. $th_{l}$ is set to 0.1 and $th_{h}$ is set to 0.4 for Cascaded Fusion.
All models are trained on 1 GPU 3090, and the batch size is 64. The learning rate is set to 0.001 while training unimodal model and 0.0001 while training HMA fusion model by the Adam optimizer. And we adopt the weighted BinaryCrossEntropy (BCE) Loss (negative:positive=5:1) to tackle the imbalance between
positive and negative samples.

\begin{table}[htb]
\small

\setlength{\tabcolsep}{1.3mm}{
  \centering
  \footnotesize
  \caption{\label{audiotable} {\it Comparison of different audio-only systems. 3D-18+2D-18 means 3D-ResNet18+2D-ResNet18.
  \vspace{+0.2em}
}}
  \begin{tabular}{ccccccccc}
  \toprule
  \multirow{2}*{\textbf{ID}} & \multirow{2}*{\textbf{Model}} & \multirow{2}*{\textbf{Field}} & \multicolumn{3}{c}{\textbf{Dev[\%]}} & \multicolumn{3}{c}{\textbf{Eval[\%]}}  \\
  \cmidrule(lr){4-6} \cmidrule(lr){7-9}  &  &  & FRR & FAR & WWS & FRR & FAR & WWS \\
  
  \midrule
  
  A1 & 2D-ResNet34 & Far & 5.45 & 4.76 & 10.21 & 5.64 & 6.05 & 11.69 \\
  A2 & 3D-ResNet34 & Far & 5.77 & 4.28 & 10.05 & 5.46 & 5.88 & 11.34 \\
  A3 & 3D-18+2D-18 & Far & 6.76 & 2.98 & 9.71 & 7.05 & 4.19 & 11.24 \\
  \midrule
  A4 & A3+SimAM & Far & 5.93 & 3.61 & 9.54 & 6.38 & 4.65 & \textbf{11.03} \\

  \bottomrule

\end{tabular}
}\vspace{-1.2em}
\end{table}

\begin{table}[htb]
\small
\setlength{\tabcolsep}{1.3mm}{
  \centering
  \footnotesize
  \caption{\label{visualtable} {\it Comparison of different video-only systems. 3D-18+2D-18 means 3D-ResNet18+2D-ResNet18.
  \vspace{+0.2em}
}}
  \begin{tabular}{ccccccccc}
  \toprule
  \multirow{2}*{\textbf{ID}} & \multirow{2}*{\textbf{Model}} & \multirow{2}*{\textbf{Field}} & \multicolumn{3}{c}{\textbf{Dev[\%]}} & \multicolumn{3}{c}{\textbf{Eval[\%]}}  \\
  \cmidrule(lr){4-6} \cmidrule(lr){7-9}  &  &  & FRR & FAR & WWS & FRR & FAR & WWS \\
  
  \midrule
  
  V1 & 3D-ResNet34 & Far & 8.81 & 8.03 & 16.85 & 18.52 & 9.03 & 27.54 \\
  V2 & 3D-18+2D-18 & Far & 9.78 & 6.64 & 16.41 & 14.41 & 9.62 & 24.03 \\
  V3 & V2+Pretrain & Far & 10.1 & 6.3 & 16.3 & 11.83 & 9.51 & 21.34 \\
  \midrule
  V4 & V3+SimAM & Far & 6.89 & 9.09 & 15.98 & 9.56 & 13.37 & 22.93 \\
  V5 & V4+Pretrain & Far & 9.13 & 6.25 & 15.39 & 8.03 & 11.1 & \textbf{19.13}  \\

  \bottomrule

\end{tabular}
}
\vspace{-1.2em}
\end{table}

\section{Results and Discussion}
\vspace{-0.5em}
\subsection{Ablation experiments for data augmentation}

We do the ablation experiments for the audio-based data augmentation to find the best combination of these methods. Table \ref{ab} reports the details results on the far-field development and evaluation set. For convenience, we use 2D-ResNet34 
for our experiments here. We find that the performance of the system degrades without any data augmentation. We first apply NR and BE to the audio-only system, and there was some improvement because of increase in the number of training samples. Based on this, SP, TS and SA
bring more significant boosts. We find that VP and NS will slightly degrade the performance of the model. However, the combination of NS and other methods will assist in performance improvement. We finally conclude the set of data augmentation methods, achieving a performance of 11.69\% WWS on the eval set.
\vspace{-0.8em}

\subsection{Results for different systems}
\label{sec:experiment}

Table \ref{audiotable} shows the results for audio-only systems of different architectures. 3D-ResNet34 performance is better than 2D-ResNet34 because of the modeling of the short temporal feature. 3D-ResNet18+2D-ResNet18 achieves better results according to the mixture of 3D and 2D CNN. Finally, the 3D-ResNet18+2D-ResNet18+SimAM achieves the best result (11.03\% WWS) without introducing extra parameters.

Table \ref{visualtable} shows the results for video-only systems of different architectures. 3D-ResNet18+2D-ResNet18 achieves better results according to the mixture of 3D and 2D CNN. The 3D-ResNet18+2D-ResNet18+SimAM achieves (22.93\% WWS) without introducing extra parameters. In addition, we follow \cite{cheng2022dku} to pretrain the backbone model on a lip-reading database named CAS-VSR-W1k database \cite{yang2019lrw}, then have it fine-tuned on the MISP database. Finally, the 3D-ResNet18+2D-ResNet18+SimAM+Pretrain achieves the best result (19.13\% WWS).

\begin{table}[htb]
\small
\setlength{\tabcolsep}{1.0mm}{
  \centering
  \footnotesize
  \caption{\label{avtable} {\it Comparison of different audio-visual systems.
  \vspace{+0.2em}
}}
  \begin{tabular}{ccccccccc}
  \toprule
  \multirow{2}*{\textbf{ID}} & \multirow{2}*{\textbf{Model}} & \multirow{2}*{\textbf{Field}} & \multicolumn{3}{c}{\textbf{Dev[\%]}} & \multicolumn{3}{c}{\textbf{Eval[\%]}}  \\
  \cmidrule(lr){4-6} \cmidrule(lr){7-9}  &  &  & FRR & FAR & WWS & FRR & FAR & WWS \\
  
  \midrule
  
  VA1 & Official \cite{zhou22g_interspeech} & Far & 7.3 & 6.8 & 14.1 & 10.1 & 15 & 25.1 \\
  VA2 & Xu et al. \cite{xu2022audio} & Far & - & - & - & - & - & 9.1 \\
  VA3 & Cheng et al. \cite{cheng2022dku} & Far & 3.85 & 3.42 & 7.27 & - & - & 7.1 \\
  VA4 & MISP 2021 1st \cite{mispchallenge}  & Far & - & - & 4.1 & - & - & 5.8 \\
  \midrule
  VA5 & Score-Level & Far & 1.92 & 3.37 & 5.29 & 1.54 & 4.82 & 6.36 \\
  VA6 & Cascaded Fusion & Far & 5.29 & 2.3 & 7.59 & 7.29 & 3.5 & 10.79 \\
  \midrule
  VA7 & HMA & Far & 3.04 & 2.55 & 5.59 & 2.15 & 3.44 & \bf{5.59} \\

  \bottomrule

\end{tabular}
}
\vspace{-1.2em}
\end{table}

The results for the audio-visual fusion systems are shown in table \ref{avtable}. Moreover, we also compare our system with the previous SOTA result. We combine the best models from the respective modalities (A4+V5) and test multiple fusion strategies. The cascade system is not good enough compared to the score-level fusion because the distribution of scores of the two modalities is inconsistent, which plays a limited screening effect and requires fine threshold adjustment. We find that Score-Level fusion and Cascaded fusion do not fully utilize the information of the visual modality. In contrast, HMA can further utilize the features of the respective modality to give multimodal results finally. And, we obtain our final model by averaging the top-3 best models which have a lower loss on the dev set. Finally, we obtain a 21\% relative reduction WSS compared with our previous work \cite{cheng2022dku} and achieve the SOTA audio-visual result (5.59\% WWS).

\vspace{-0.1em}
\section{Conclusion}
\vspace{-0.5em}
In this paper, we extend our previous work \cite{cheng2022dku} to investigate robust audio-visual WWS system. We use a hybrid 3D and 2D convolution network to model the low-level spatial and high-level semantic information, respectively, while introducing the SimAM without additional parameters to improve the performance of the unimodal network further. We also explore different multimodal fusion schemes and find that HMA can take full advantage of the information from both modalities to give the best results. The performance of the system has been significantly improved compared with our previous work and achieved SOTA result (5.59 \% WWS) on the MISP dataset. 

\vspace{-0.1em}
\section{ACKNOWLEDGMENTS}
\label{sec:copyright}
\vspace{-0.5em}

This research is funded in part by the National Natural Science Foundation of China (62171207), Science and Technology Program of Guangzhou City (202007030011), Science and Technology Program of Suzhou City(SYC2022051), and Alibaba Air Fund. Many thanks for the computational resource provided by the Advanced Computing East China Sub-Center.




\bibliographystyle{IEEEbib}
\bibliography{strings,refs}

\end{document}